\documentclass[journal]{IEEEtran}

\usepackage{tabularx}
\usepackage[tight,footnotesize]{subfigure}
\usepackage{epsfig, cite, rotating}
\usepackage[cmex10]{amsmath}

\usepackage{graphicx}
\graphicspath{{figures/}}

\DeclareMathOperator*{\minimize}{Minimize}
\DeclareMathOperator*{\argmax}{argmax}
\DeclareMathOperator*{\argmin}{argmin}

\title{Correlation-aware Resource Allocation in Multi-Cell Networks}

\author{\IEEEauthorblockN{Dorna Bandari\IEEEauthorrefmark{1},
Gregory Pottie\IEEEauthorrefmark{1},
 Pascal Frossard\IEEEauthorrefmark{2}}
\IEEEauthorblockA{\IEEEauthorrefmark{1}University of California, Los Angeles, USA\\}
\IEEEauthorblockA{\IEEEauthorrefmark{2}\'{E}cole Polytechnique F\'{e}d\'{e}rale de Lausanne (EPFL), Switzerland\\
Email:  dorna@ucla.edu,  pottie@icsl.ucla.edu, pascal.frossard@epfl.ch}}

\author{Dorna~Bandari,~\IEEEmembership{Member,~IEEE,}
        Gregory~Pottie,~\IEEEmembership{Fellow,~IEEE,}
        and~Pascal~Frossard,~\IEEEmembership{Senior Member,~IEEE}
\thanks{D. Bandari is with the Department
of Electrical Engineering, University of California, Los Angeles, CA, 90025 USA e-mail: dorna@ucla.edu.}
\thanks{G. Pottie is with  University of California, Los Angeles and P. Frossard is with \'{E}cole Polytechnique F\'{e}d\'{e}rale de Lausanne (EPFL), Switzerland.}}

\begin{document}
\maketitle
\begin{abstract}

We propose a cross-layer strategy for resource allocation between spatially correlated sources in the uplink of multi-cell FDMA networks. Our objective is to find the optimum power and channel to sources, in order to minimize the maximum distortion achieved by any source in the network. 
Given that the network is multi-cell, the inter-cell interference must also be taken into consideration.
This resource allocation problem is NP-hard and the optimal solution can only be found by exhaustive search over the entire solution space, which is not computationally feasible. We propose a three step method to be performed separately by the scheduler in each cell, which finds cross-layer resource allocation in simple steps. The three-step algorithm separates the problem into inter-cell resource management, grouping of sources for joint decoding, and intra-cell channel assignment. For each of the steps we propose allocation methods that satisfy different design constraints. 
In the simulations we compare methods for each step of the algorithm. We also demonstrate the overall gain of using correlation-aware resource allocation for a typical multi-cell network of Gaussian sources. We show that, while using correlation in compression and joint decoding can achieve 25\% loss in distortion over independent decoding, this loss can be increased to 37\% when correlation is also utilized in resource allocation method. This significant distortion loss motivates further work in correlation-aware resource allocation. Overall, we find that our method achieves a 60\% decrease in 5 percentile distortion compared to independent methods.
\end{abstract}

\section{Introduction}
\begin{figure}[htb]
\begin{minipage}[b]{1.0\linewidth}
 \centering
 \centerline{\epsfig{figure=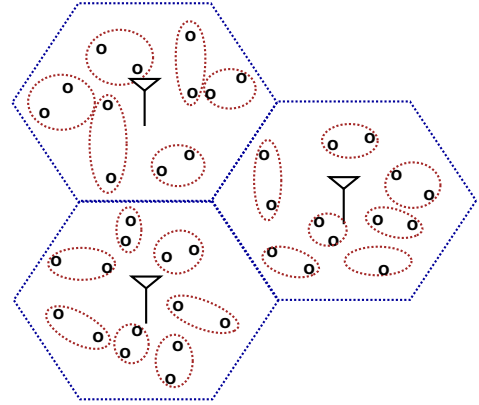,width=8.5cm, trim=2cm 14cm 3cm 3.5cm}} % L B R Top
 \caption{Three neighboring cells with base stations in the center of each cell. The sensors are grouped in correlated groups of size 2 in this example.}
\label{fig:systemcells}
\end{minipage}
\end{figure}

We examine a scenario where spatially correlated sources in a multi-cell Frequency Division Multiple Access (FDMA) network transmit data to the base station of their cell. An example is shown in Figure \ref{fig:systemcells}, where the sources are scattered on a field that is divided into three hexagonal cells. The scenario we consider in this work readily applies to Wireless Sensor Networks (WSNs), where sensors measure various spatially correlated phenomena such as temperature, humidity, audio, video, etc. \cite{Akyildiz2002}. Since much of the surface area of interest for WSN applications is now covered by cellular networks, it is more accessible for some WSNs to simply use this existing communication infrastructure. Therefore in this work we assume a multi cell network with a Medium Access Control (MAC) scheme similar to LTE of 3G and WiMAX.

The base station in the center of each cell contains a scheduler that performs resource allocation with the aim of minimizing the maximum individual distortion in the global network after data reconstruction. Our goal is to find a strategy to efficiently allocate resources to each source, while taking advantage of spatial correlation among the sources.

The resource allocation problem for FDMA networks involves finding the channel and power assignment for sources in each cell, given the benefits and costs of this assignment  (i.e., utility gain versus the interference to others). Even when sources are independent, optimal allocation of power and channels in an interfering network of users is a non-convex, mixed integer non-linear programming problem (MINLP). The independent allocation problem is strongly NP-hard when there are several channels to choose from \cite{Luo08}. Furthermore, for a non-trivial network size exhaustive solutions are not computationally feasible in a single scheduling period. Therefore further simplifications and approximations are needed. Adding correlation to the design only makes this problem more complex. 
The correlation characteristics of sources and their contribution to the performance of other sources in the network should be considered in the resource allocation scheme. The overall resource allocation problem becomes computationally complex and simplifications are necessary in practice towards effective (but often suboptimal) solutions.

In this work we present a three-step approach for the problem of resource allocation between spatially correlated sources. Namely, the three steps are called \emph{i) Inter-cell resource management, ii) Source Grouping,} and \emph{iii) Intra-cell Scheduling}. Separating the overall resource allocation problem into three sub-problems leads to an efficient solution with tractable complexity.

First, we find the transmit power limits in each cell in the inter cell interference management step. For this, we use the Adaptive ICon interference coordination method \cite{Wimob2011}. ICon manages the inter-cell interference by concentrating the interference experienced by a cell base station in a designated frequency sub-band. This results in setting transmit power limits on each sub-band for the users of neighboring cells, and separates the global problem into smaller problems solvable in each cell, while adaptively managing the inter-cell interference. Adaptation is performed in order to balance the performance across the network. Second, in the Source Grouping (SG) step we split the set of sources in each cell to smaller groups to be jointly decoded. This is done in order to take advantage of spatial correlation without high increase in complexity in the decoder and scheduler. We compare the utility gain of larger group sizes and conclude that having two sources per group achieves most of the correlation gain. We propose a location-based adaptive SG method called Distance Outer-Priority (distance OP). This method finds the best grouping for the cell, while accounting for the interference to neighboring cells. Last, we use the parameters found in previous steps in order to allocate channels in the Intra-cell scheduling step using two novel scheduling methods. One scheme called Distortion Proportionally Fair (D-PF) is based on the popular Proportionally Fair method \cite{Wengerter2005}, modified in order to take the effects of correlation into account. The second scheduling method is a linear programming problem (OPT), which performs better than D-PF at the price of higher complexity. 

We finally demonstrate in simulation that our proposed scheme presents a constructive, simple solution to the computationally infeasible optimum resource allocation problem. The rate-distortion performance is superior to resource allocation methods that do not consider correlation in the optimization. It confirms the potential of the proposed solution for data gathering in wireless sensor networks. 

The paper is organized as follows. In Section \ref{sec:relwork} we discuss the related work. We formulate the problem in Section \ref{sec:probform}, and propose our three phased solution in Section \ref{sec:solution}. We present the results of our simulations in Section \ref{sec:sim}.

\section{Related Work}
\label{sec:relwork}

Resource allocation in this work refers to finding joint power and channel allocation to multiple sources in the same cellular network. 
For this problem, common approaches are based on decoupling the inter-cell and intra-cell resource management problems.
For inter-cell resource management, Fractional Frequency Reuse (FFR) and Soft Frequency Reuse (SFR) \cite{Halpern1983} \cite{Huawei2005} have been proposed by the cellular industry. In these methods, a \emph{power profile} is assigned to every cell, which sets transmit power limits on each frequency sub-band for Orthogonal FDMA (OFDMA). The neighboring cells are assigned complementary power profiles in order to shape the inter-cell interference. These power profiles are based on heuristics. For intra-cell scheduling, the scheduler in each cell usually performs channel allocation independently using a method such as Proportionally Fair scheduling \cite{Wengerter2005}. Although these methods are promising towards simultaneous power and channel allocation, their performance is not better in general than the frequency Reuse 1 scheme \cite{Elayoubi2008}, which simply permits the use of every frequency band at the same transmit power limit in every cell. 

In order to improve on these solutions, we propose Adaptive ICon in \cite{Wimob2011}. Adaptive ICon is an adaptive inter-cell interference management method based on concentrating the interference to a cell on a particular sub-band. This is achieved by requiring the neighboring cells to respect an \emph{interference power profile} set by the cell. The method achieves significant improvement over FFR and SFR schemes. Another interesting inter-cell interference management method called \emph{inverted reuse} is proposed in \cite{gerlach2010}; it is also based on concentrating the interference experienced by a cell on a specific frequency band by setting transmit power limits for its neighboring cells. 

On the other hand, the consideration of spatial correlation in resource allocation has been studied mainly in the field of WSNs. Early methods use 802.11-like MAC, and save power by enabling periodic sleep-wake cycles for wireless sensors, with algorithms such as S-MAC \cite{Ye2002} and P-MAC\cite{Radhakrishnan2003}. Later the authors in \cite{Vuran2006} proposed Correlation-based Collaborative Medium Access Control (CC-MAC), which takes advantage of spatial correlation among sensors by finding a subset of sensors to transmit their data while data from the other sensors is omitted. In this work we rather use the data from every sensor, while minimizing the maximum distortion in the network. 

A different but related problem is studied in multi-hop WSNs, where spatial correlation is utilized in combining routing and rate allocation with data compression \cite{Pattem2004} \cite{Li2009} \cite{Wang2009} \cite{Moad2011}. In \cite{Cristescu2006}, the authors propose a method for lossy transmission of data, combining routing with rate allocation using Wyner-Ziv (WZ) distributed source coding \cite{Wyner1976}. They find that routing and rate optimization can be decoupled and optimized separately, when the cost function is a weighted sum of rates. Although this is a different problem than the one we are considering in this work, the ideas developed in this area are similar to our approach to the joint power and channel allocation problem for correlated sources. In particular, correlation is exploited by a joint decoder in order to reduce the overall resource utilization. To the best of our knowledge there is however no work that proposes solutions for the joint power and channel allocation for correlated sources in multi-cell networks.

\section{Resource Allocation Problem}
\label{sec:probform}

We consider a 2-D multi-cell network with FDMA multiple access scheme. The sources sense spatially correlated information. They transmit their observations to the base station of their cell, using medium access parameters determined by the scheduler located in the cell base station. 

\subsection{Resources}
In FDMA the bandwidth is split into frequency sub-bands and a subset of these sub-bands is assigned to each user. In the multi-cell scenario, users from different cells interfere if they are assigned to the same channel. Therefore, the transmission power per channel for each user is an important factor in network performance. The MAC problem consists in determining the power per channel and channel allocation among users. Each of the independent orthogonal channels are assumed to be Gaussian and interference is considered to be equivalent to noise with respect to channel capacity. 
The following rate is achieved for source \(s_i\) in cell \emph{k} according to Gaussian channel capacity \cite{Cover1991}:
\begin{align}
R_{i} = \sum_{c=1}^{C} a_{i,c}B_{c}\log_2\left(1+ { p_{i,c}.g_{i,k}\over N_0.BW_{c}+ \sum_{u \in U_k} P^I_{ukc} }\right) % \forall i \in X
\label{eq:rate}
\end{align} 
where \( a_{i,c}\) is the binary value specifying channel allocation and \(p_{i,c}\) the transmit power of source \(s_{i}\) on channel \emph{c}, \(g_{i,k}\) is the channel gain from source \(s_{i}\) to receiver of cell \emph{k}, and \(B_{c}\) is the bandwidth of channel \emph{c}. \(N_0\) is the noise power, \(U_k\) is the set of neighboring cells of cell \(k\). \( P^I_{ukc}\) is the received interference power from cell \emph{u} to cell \emph{k} on channel \emph{c}; namely, \( P^I_{ukc} = \sum_{s_j \in X_u} a_{j,c}.p_{j,c}.g_{j,k} \), where \(X_u\) is the set of all the sources in cell \(u\).

\subsection{Rate-Distortion Region}
Now we must relate the rate available to a source to the distortion after reconstruction. This is achieved by rate-distortion (R-D) function, which is a characteristic of the coding scheme. For general lossy distributed coding, the R-D region is not yet known.  However, for distributed coding in the limit of \emph{high resolution}, i.e., as the quantization resolution increases, the R-D region is known to be similar to the rate region given by lossless Slepian-Wolf coding \cite{Berger99}. When sources \(\boldsymbol S\) in set \(\boldsymbol G\) in cell \emph{k} are decoded jointly, the R-D region for this set in the limit of high resolution is given by:
\begin{align}
& \forall \boldsymbol S \subseteq \boldsymbol G : \nonumber \\
& \sum_{s_i \in \boldsymbol S}R_{i} \geq h_2(\boldsymbol S | \boldsymbol G \backslash \boldsymbol S) -{ \frac 1 2} \log_2 \left((2\pi e)^{|\boldsymbol S|} \prod_{s_i \in \boldsymbol S} D_{i} \right) 
\label{eq:rd}
\end{align}
where \(h(.)\) is the differential entropy, and \(D_{i}\) is the squared error distortion of source \(s_{i}\). This is an outer bound for the general coding case; it becomes tighter as the resolution increases, thus increasing the accuracy of the model.

In order to find the entropies in Eq. (\ref{eq:rd}) we model the sources as joint random variables with some known distribution, with a distance based correlation model. The observations at the sources are modeled as joint Gaussian random variables, which is commonly used in WSNs. For the correlation model we use an exponential distance based correlation model \cite{Berger2001}. The entropy is then given by,
\begin{align}
h_2(\boldsymbol S) = {\frac 1 2} \log_2\left( (2 \pi e)^{|\boldsymbol S|} |\Sigma|\right) 
\label{eq:ent}
\end{align}
where \(|\Sigma|\) is the determinant of the covariance matrix with elements given as below, when \(\sigma^2\) is the variance of the sources:
\begin{align}
\sigma^2_{ij} = \sigma^2 .e^{({\frac {-d_{ij}} \theta})}
\label{eq:corr}
\end{align}
Where \(d_{ij}\) is the distance between sources \(s_{i}\) and \(s_{j}\), and \(\theta\) is the correlation model parameter. We use the model in Eq. (\ref{eq:rd}) in our resource allocation framework, but it can be replaced without significantly changing our resource allocation methodology. Additionally, the choice of the distribution and the correlation model does not affect our analysis. The only requirement for the correlation model is that correlation must decrease with distance. 

The set of sources, \(\boldsymbol G\), that are decoded jointly can include between 1 source and all the sources in a cell. Increasing the size of joint decoding groups increases the complexity in decoding and scheduling. Also, we later show that increasing the group size has diminishing returns in terms of decrease in distortion. Therefore we assume that sources are decoded in small groups.

\subsection{Problem Formulation}
The optimization problem is to minimize the maximum distortion, subject to resource constraints. Assuming there are \(N\) sources in the network and \(X_k\) is the set of all sources in cell \emph{k}, the resource allocation problem is given as follows.

\begin{align}
\text{Problem 1:}\nonumber \\
\minimize_{\boldsymbol  {a},\boldsymbol p, \boldsymbol G} \:\:\:  &\max_{i} (D_{ i})  \\
s.t.\:\:\:\:\: &\sum_{s_i \in \boldsymbol S}R_{i} \geq -{ \frac 1 2} \log_2 \left((2\pi e)^{|\boldsymbol S|} \prod_{s_i \in \boldsymbol S} D_{i} \right)  \nonumber \\
&\:\:\:\:\:\:+ h_2(\boldsymbol S | \boldsymbol G^j_{k}\backslash \boldsymbol S),\:\:\:\:\: \:\:\:\:\:\:\:\:\:\: \: \forall \boldsymbol S \subseteq \boldsymbol G^j_{k}, \forall j, \boldsymbol G_{k} , k \nonumber \\
&\sum_{s_i \in X_k} a_{i,c} = 1, \:\:\:\:\:\:\:\:\:\:\: \:\:\: \: \:\:\:\:\:\:\:\:\:\:\: \:\:\:\:\:\:\:\:\:\:\:  \:\:\:\:\:\:\:\:\:\:\:\:\:\:\:\:\: \forall c,k \nonumber \\
&P_{MIN} \leq p_{i,c} \leq  P_{MAX},\:\:\:\:\:\:\:\:\:\:\:\:\:\:\:\:\:\:\:\:\:\:\:\:\:\:\:\:\:\: \:\:\:\:\:\:\:\:\: \forall i, c \nonumber
\label{eq:overallopt}
\end{align}
where \(R_{i}\) is given in Eq. (\ref{eq:rate}). \(P_{MIN}\) and \(P_{MAX}\) are the transmit power limits.  \(\boldsymbol {G}_k\) is the source grouping, i.e., set of all the correlated sets in cell k, and \(\boldsymbol G^j_{k}\) is the \(j^{th}\) correlated set in this cell. Parameter \(\boldsymbol S\) is a possible subset of the correlated set \(\boldsymbol G^j_{k}\). 

The aim is to find the optimum resource allocation vectors \(\boldsymbol  {a}^*\) and \(\boldsymbol p^*\); and the source grouping, \(\boldsymbol {G}^* = [\boldsymbol G^*_1, ..., \boldsymbol G^*_K]\). This problem is NP-hard for \(C>1\) based on similar arguments as those given in \cite{Luo08}. We therefore propose in the next section to decompose this problem into three smaller problems that can be solved efficiently.

\section{Three-Step Solution}
\label{sec:solution}

\begin{figure}[htb]
\begin{minipage}[b]{1.0\linewidth}
 \centering
 \centerline{\epsfig{figure=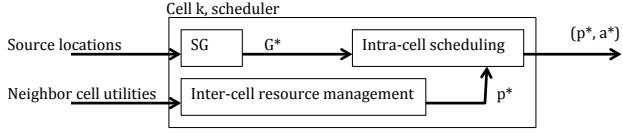,width=8.5cm, trim=2cm 17cm 3cm 7cm}} % L B R Top
 \caption{Scheduler in the base stations.}
\label{fig:sched}
\end{minipage}
\end{figure}

We separate the resource allocation problem into three steps, as demonstrated in Figure \ref{fig:sched}. The three steps are performed in the scheduler, located in the base station of every cell. The steps are called Inter-cell resource management, Source Grouping (SG), and Intra-cell Scheduling. 

\subsection {Approximate Solution}

The Inter-cell resource management finds the transmit power limits for each user on each sub-band, given the location of the sources on the field and performance of the neighboring cells. We use Adaptive ICon \cite{Wimob2011} for this step, which is based on concentrating the interference to a cell on a portion of the bandwidth. The ICon parameters are adapted in order to balance the utility (maximum distortion in this problem) across the cellular network by increasing or decreasing interference limits of a cell depending on its utility relative to its neighbors. The utility achieved in a cell and its neighbors is communicated among the base-stations. 

In the Source Grouping (SG) step, the sources are grouped together for joint decoding (which is then used in the intra-cell scheduling step). We examine having correlated groups of one, two, and three sources. We propose a distance-based SG, which also takes the potential added interference of the sources into consideration when finding correlated groups.

The intra-cell scheduling step involves allocating channels to users, given the power and source grouping found in the previous steps. Given that schedulers can have different computational capacity, we propose two methods with different complexities/benefits: 1) a very simple method called D-PF, similar to the common Proportionally Fair (PF) scheduling \cite{Wengerter2005}, modified to take correlation into account, 2) a more complex, but still polynomial time solvable linear programming method, which is a relaxed version of the integer programming scheduling problem. 

Each step is performed at different time scales, depending on the design choices. Typically the inter-cell resource management is performed infrequently, since it requires feedback from the decoder. Source grouping is performed whenever sources are moved, or once nodes are added or removed. The intra-cell scheduling is performed often, with a frequency depending on the choice (and therefore complexity) of the scheduling algorithm. 

\subsection {Inter-Cell Resource Management}
\label{sec:inter}
In this step, we find a rule for allocating resources among interfering cells. If no rule is chosen, the Inter-cell resource management is effectively a Reuse 1 scheme, i.e. all resources are used in all cells. FFR and SFR could also be used in this step. We propose to use Adaptive ICon \cite{Wimob2011}, which improves on the Reuse 1, FFR, and SFR schemes for the worst performing users, as we will demonstrate in the simulations.

\begin{figure}[htb]
\begin{minipage}[b]{1.0\linewidth}
 \centering
 \centerline{\epsfig{figure=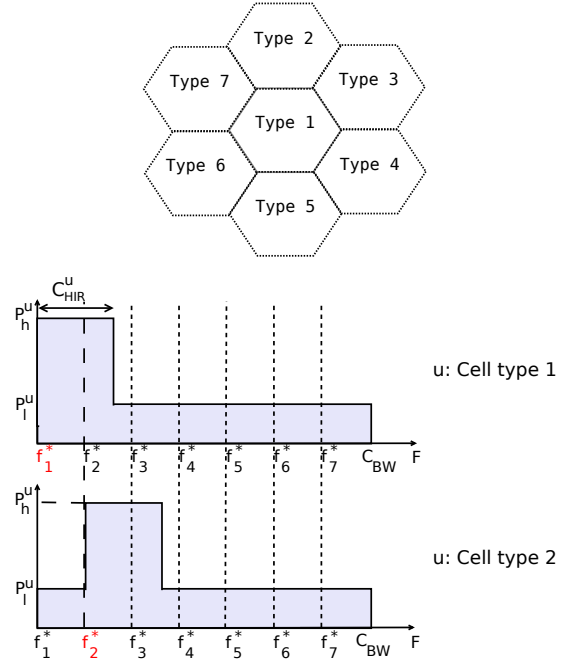,width=12cm, trim=1cm 13.5cm 5cm 3cm}} % L B R Top
 \caption{Examples of Interference Power Profiles (IPPs) set by ICon method for cell types 1 and 2 in a hexagonal cellular network. \(\text{f}^*_i\)s are the starting sub-bands for the High Interference Region (HIR) of each cell type. \(C^u_{HIR}\) is the number of sub-bands in HIR, and can be any value between 0 and \(C_{BW}\).}
\label{fig:ICon}
\end{minipage}
\end{figure}

ICon is an inter-cell resource management method based on defining Interference Power Profiles (IPPs). The IPP defines a limit to received interference on each frequency sub-band, which the neighboring cells are obliged to meet. An example is given in Figure \ref{fig:ICon}. This is a heuristics-based method, based on the following intuition: if the strong interferers in all neighboring cells are concentrated on the same sub-band, the bandwidth will be used more efficiently. The design parameters are \(P_h^u\) and \(C^u_{HIR}\), respectively the maximum received interference and width of the \emph{high interference region} for cell \emph{u}. We assume that these parameters are known for a given network structure, and are later adapted in order to balance the interference across the cells. Adaptation of ICon can be performed efficiently, with minimal inter-cell communication. 

\begin{figure}[htb]
\begin{minipage}[b]{1.0\linewidth}
 \centering
 \centerline{\epsfig{figure=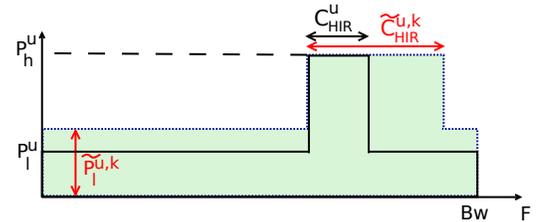,width=12cm, trim=1cm 21cm 7cm 4cm}} % L B R Top
 \caption{Interference Power Profile (IPP) for cell \emph{u}. Green curve is the IPP that cell \emph{k} has to comply with towards cell \emph{u}. }
\label{fig:IPPbasedAdap}
\end{minipage}
\end{figure}

We initially set all the IPPs equal to the IPP found for the given network structure for the particular cell type. Example IPP are shown in Figure \ref{fig:IPPbasedAdap}. Their attributes are modified by the following adaptation strategy. At each Inter-cell adaptation step, each base station sends the utility achieved by in the pervious period to its neighbors. Then each cell updates the the IPP it imposes on each of its neighbors. This way, two neighbors with different utilities are assigned different IPPs. The new neighbor-specific parameters are given as follows for IPP of cell \emph{k} with neighbor \emph{u},
\begin{align}
&\widetilde{C}^{u,k}_{HIR} (t+1) = \max \Bigl\{\widetilde{C}^{u,k}_{HIR} (t) -\nonumber \\
& \:\:\:\:\:\:\:\:\:\:\:\:\:\:\:\:\:\:\:\:\:\:\:\:\:\:\:\:\:\: \alpha. \left( \frac{ U^u(t) -U^k(t)} {(U^u(t) +U^k(t))/2}\right). C_{Bw} , C^u_{HIR}  \Bigr\} \nonumber
\end{align}
and the low interference power limit, \( \widetilde{P}_l^{u}\) is updated as,
\begin{align}
&\widetilde{P}_l^{u,k}(t+1) = \max \Bigl\{\widetilde{P}_l^{u,k}(t) - \nonumber \\
&\:\:\:\:\:\:\:\:\:\:\:\:\:\:\:\:\:\:\:\:\:\:\:\:\:\:\:\:\:\: \beta.  \left( \frac{ U^u(t) -U^k(t)} {(U^u(t) +U^k(t))/2}\right).  P^u_h  , {P}_l^{u}  \Bigr\} \nonumber
\end{align}
where \(U^k(t)\) is the utility achieved in cell \(k\) at time \(t\). For this problem,  \(U^k(t) = \max_{i} (D_{i,k})\). The parameters \(\alpha \text{ and } \beta\) are step size values in the range \([0, 1]\). The above updates should be transmitted to the base station of each neighboring cell.

The definition of IPP in neighboring cells sets power limits for each of the sources, in order to avoid interference. Namely, for each user \emph{i} in cell \emph{k} and each channel \emph{c}, the following inequalities must hold:
\begin{align}
&p_{i,c}.{g_{i,u}} \leq {I_u (c)} & \forall u
\end{align}
where \(I_u (c)\) is the value of IPP of cell \emph{u} at channel \emph{c}, and \({g_{i,u}}\) is the channel gain from user \emph{i} to base station of cell \emph{u}. This value is known at the user if we assume channel reciprocity and can be transmitted to the base station of cell \emph{k}. The maximum transmit power that does not violate any of the neighbors' IPP is thus determined as
\begin{align}
p^{max}_{i,c} = \min_{u, u\neq k} {\frac {I_u (c)}  {g_{i,u}}}.
\end{align}
These transmit power limits set by the inter-cell resource management step control and limit the interference in the network. Namely, a cell scheduler respecting these limits can perform its scheduling independently, without penalizing the other cells. 

Going back to the optimization problem given in Eq. (\ref{eq:overallopt}) and separating the problem to be solved independently in each cell, we find that the objective function in each cell is decreasing in \( \boldsymbol p\). In other words, if assigned channel \emph{c}, user \emph{i} should transmit at maximum power within its own transmit power limits,
\begin{align}
p^*_{i,c} = max(min( p^{max}_{i,c}, P_{MAX} ), P_{MIN})
\label{eq:transmitP}
\end{align}
In other words, no other power assignment can achieve higher utility in cell \emph{k} within the constraints of the problem. We thus know the maximum rate for each user on each channel. From Eq. (\ref{eq:rate}), we can write the rate of user \emph{i} on channel \emph{c} as
\begin{align}
R^*_{i,c} = BW_{c}\log_2\left(1+ { p^*_{i,c}.g_{i,k}\over N_0.BW_{c}+ P^I_{kc}}\right) % \forall i \in X
\label{eq:possrate}
\end{align} 
where \( P^I_{kc}\) is the total interference that the base station of cell \emph{k} measures on channel \emph{c}. We can update the optimization problem of Eq. (\ref{eq:overallopt}) by fixing the power, and thus the rate per channel, to the value defined by inter-cell resource management. In every cell \emph{k} we have,
\begin{align}
\text{Problem 2:}\nonumber\\
\minimize_{\boldsymbol  {a}, \boldsymbol  {G}_k} \:\:\:  &\max_{s_i \in X_k} (D_{i}) \label{eq:indepopt} \\ 
s.t.\:\:\:\:\: &\sum_{s_i \in \boldsymbol S}\sum_{c =1}^C  a_{i,c} R^*_{i,c} \geq -{ \frac 1 2} \log_2 \left((2\pi e)^{|\boldsymbol S|} \prod_{s_i \in \boldsymbol S} D_{i} \right)  \nonumber \\
& \:\:\:\:\:\:\:\:\:\:\: \:\:\:\: +h_2(\boldsymbol S | \boldsymbol G^j_{k}\backslash \boldsymbol S), \:\:\:\:\:\:\:\:\:\:\:\: \forall \boldsymbol S \subseteq \boldsymbol G^j_{k}, \forall j, \boldsymbol G_{k} \nonumber \\
&\sum_{s_i \in X_k} a_{i,c} = 1 \:\:\:\:\:\:\:\:\:\:\: \:\:\:\:\: \:\:\:\:\: \:\:\:\:\:\:\:\:\:\:\: \:\:\:\:\:\:  \:\:\:\:\:\:\:\:\:\:\:\:\:\:\:\:\:\:\:\:\:\: \forall c \nonumber
\end{align}
The variables now are the channel allocation matrix \( \boldsymbol a^*\) and the optimum grouping \( \boldsymbol G^* \). In the next step we find \( \boldsymbol G^*\), which simplifies the scheduling further.

\subsection{Grouping of Correlated Sources}
\label{sec:SG}
\begin{figure*}[!t]
\centerline{\subfigure[2 per set, distance-based method without OP.]{\includegraphics[width=5.5cm, trim=3cm 16.5cm 5cm 3cm]{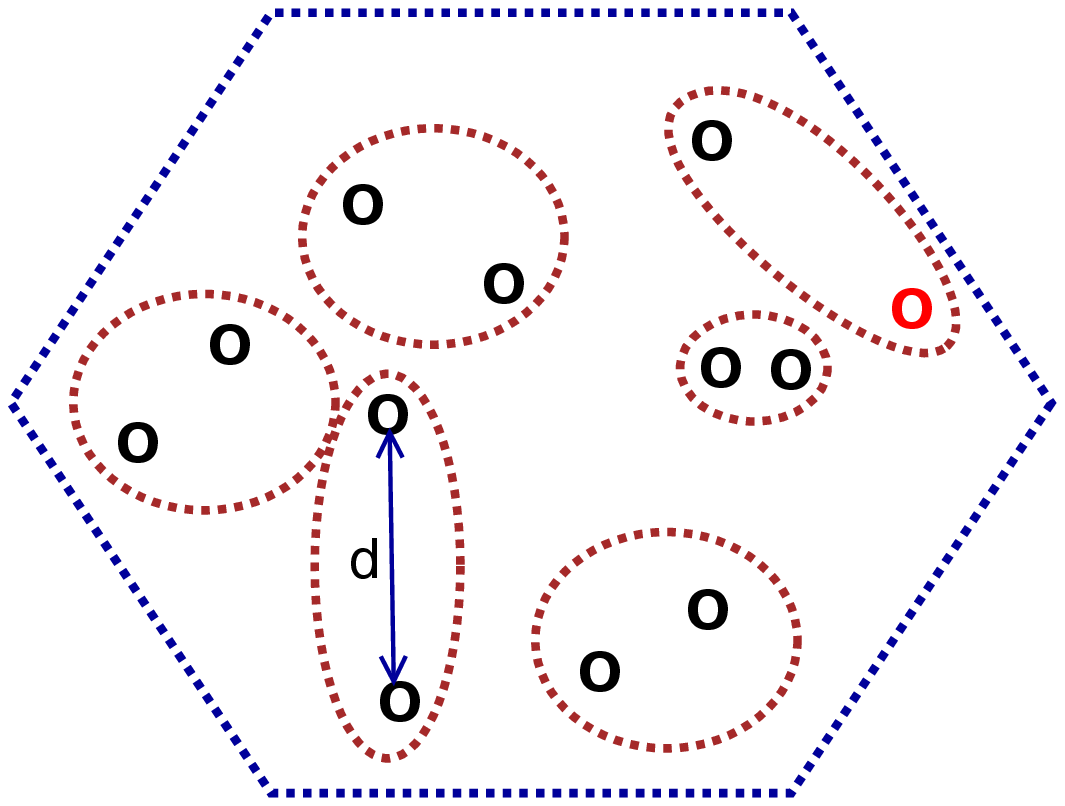} % L B R Top
\label{fig:groupingdistance}}
{\subfigure[2 per set, distance-based method with OP.]{\includegraphics[width=5.5cm, trim=3cm 16.5cm 5cm 3cm]{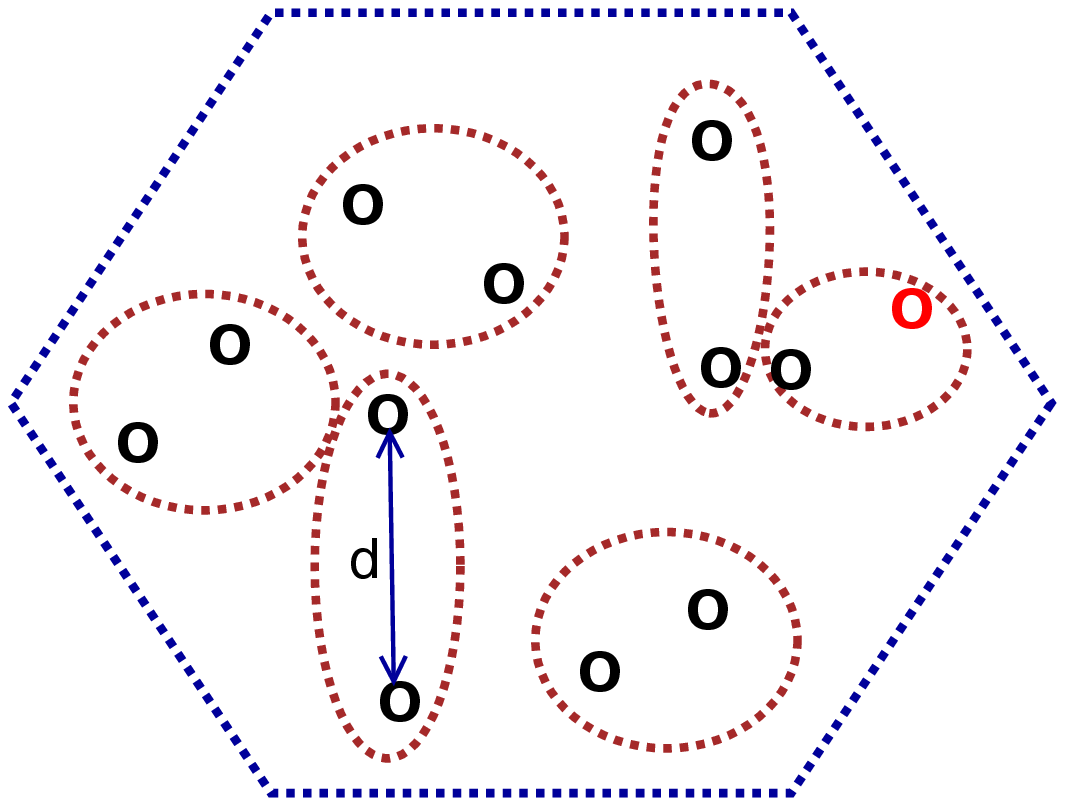}
\label{fig:outerpriority}}
{\subfigure[3 per set, distance-based method with OP]{\includegraphics[width=5.5cm, trim=3cm 16.5cm 5cm 3cm]{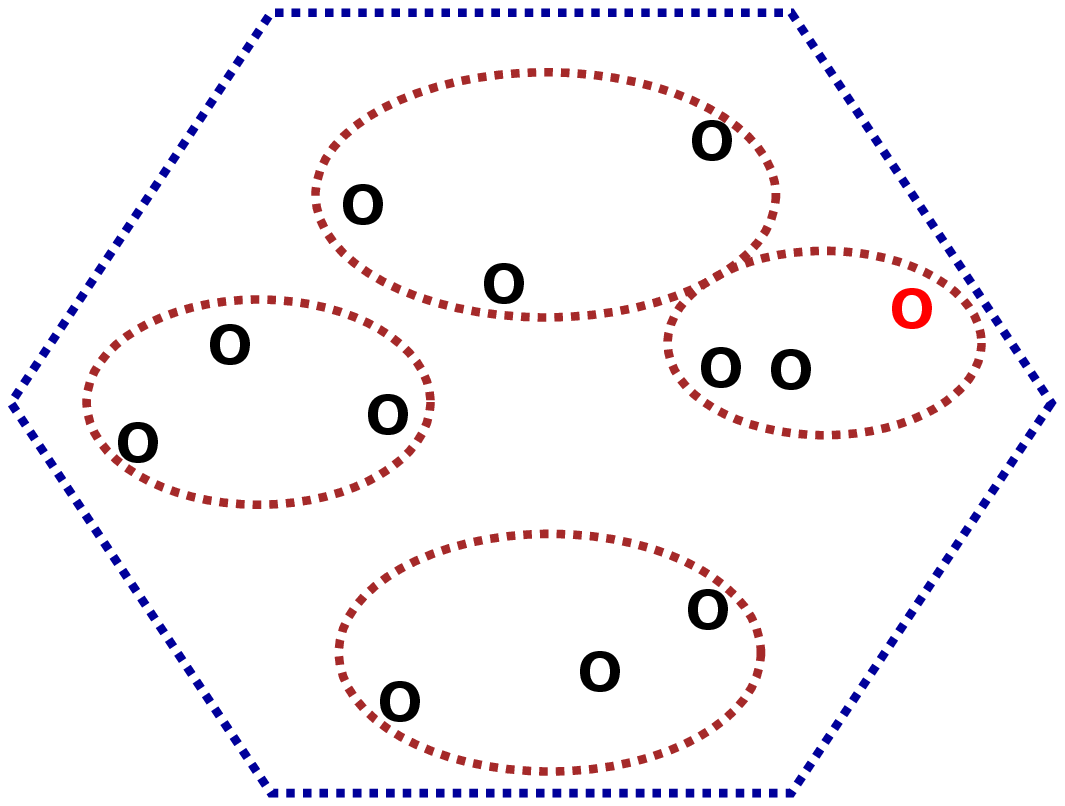}
\label{fig:outerpriority3pset}}}}}
\caption{Example of output of grouping algorithms. (a) result of the distance based grouping method without outer priority (OP); (b) groups resulting from distance based method with OP. The effects of OP can be seen by observing the change of grouping that occurs around the source drawn in red; (c) grouping results with distance-based method with OP for groups of three sources.}
\label{fig:groupings}
\end{figure*}

In order to take advantage of correlation in our framework where sources are encoded independently, we need to make use of joint decoding. The number of sources that are jointly decoded can theoretically be as many as all the sources in a cell with correlation benefits increasing with the number of jointly decoded sources. However, increasing the group size beyond two or three sources per group does not offer significant benefits in terms of distortion. Additionally, complexities of both decoding and cross layer scheduling increase with the number of sources involved in joint decoding. Therefore in this work we assume that only a small number of sources can be decoded jointly, i.e. two or three. Initially, we assume that the size of the groups is set and fixed. The scheduler has now find which sources should be grouped together and jointly decoded in order to maximize the benefits.

Grouping affects the performance in two ways, directly and indirectly. The direct effect refers to the performance a cell achieves as a result of jointly decoding the given groups of sources together. Specifically, the correlation levels of sources that are grouped together and the channel rate that each source achieves affect the utility in the cell. On the other hand, the indirect effect refers to the impact of the particular grouping method on the cross-layer resource allocation strategy, and therefore on the interference levels in the network, which results in change in the utility achieved in each cell.

The direct effect of a particular grouping of sources can be estimated using the rate-distortion region given in Eq. (\ref{eq:rd}), if the data rates of the sources are fixed. In order to simplify the analysis, we demonstrate this effect for a group of two correlated sources. The joint rate-distortion region is given by the following three inequalities:
\begin{figure}[htb]
\begin{minipage}[b]{1.0\linewidth}
 \centering
 \centerline{\epsfig{figure=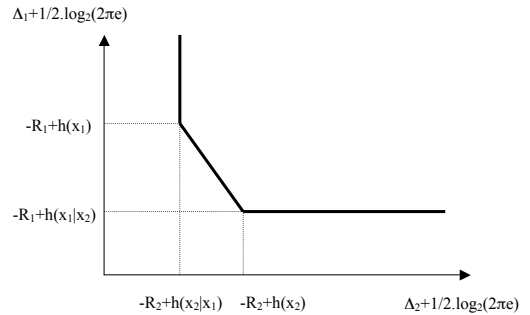,width=8.5cm, trim=4cm 18cm 4cm 4cm}} % L B R Top
 \caption{Trade-off of distortions for two correlated users \cite{Berger99}. }
\label{fig:RDcurve}
\end{minipage}
\end{figure}
\begin{align}
&R_1 \geq h_2(X_1|X_2) -{ \frac 1 2} \log_2 \left(2\pi e D_1 \right)  \nonumber\\
&R_2 \geq h_2(X_2|X_1) -{ \frac 1 2} \log_2 \left(2\pi e D_2 \right) \nonumber\\
&R_1+R_2\geq h_2(X1,X_2 ) -{ \frac 1 2} \log_2 \left((2\pi e)^2 D_1D_2 \right) \nonumber
\end{align}
We define \(\Delta_i = 1/2 \log_2 D_i\) to simplify the notation. It is bounded by:
\begin{align}
&\Delta_1 \geq -R_1+h_2(X_1|X_2) -{ \frac 1 2} \log_2 (2\pi e ) \label{eq:tradeoff} \\
&\Delta_2 \geq -R_2+h_2(X_2|X_1) -{ \frac 1 2} \log_2 (2\pi e ) \nonumber \\
&\Delta_1+\Delta_2\geq  -R_1 -R_2+h_2(X1,X_2 ) -\log_2 (2\pi e) \nonumber
\end{align}
The corresponding region is illustrated in Figure \ref{fig:RDcurve}. When the aim is to minimize the maximum distortion, the following conditions determine the optimal distortion among the users in a correlated group:
\begin{align}
&\text{if}  \:\:\: -R_i+h_2(X_i|X_j) \geq  -R_j+h_2(X_j) \label{eq:ld} \\
&\:\:\:\:\:\:\:\:\:\:\:\:\:\:\:\:\:\:\:\:\:\:\:\:\:\: \text{ then } \Delta_i =-R_i+h_2(X_i|X_j) -{ \frac 1 2} \log_2 (2\pi e ) \nonumber\\
&\:\:\:\:\:\:\:\:\:\:\:\:\:\:\:\:\:\:\:\:\:\:\:\:\:\:\text{ and } \; \Delta_j = -R_i+h_2(X_j) -{ \frac 1 2} \log_2 (2\pi e ) \nonumber
\end{align}
for i=1, j = 2, and vice versa.  If neither condition is met, then the distortions will be equal:
\begin{align}
&\Delta_1 = \frac 1 2 \left[ -R_1 -R_2+h_2(X1,X_2 ) -\log_2 (2\pi e) \right]   \nonumber \\
&\Delta_2 = \Delta_1.
\label{eq:ldeq} 
\end{align}
Conditions of Eqs (\ref{eq:ld}) and (\ref{eq:ldeq}) can be directly observed in Figure \ref{fig:RDcurve}. Note that in each of the above conditions one of the three inequalities given in Eq. (\ref{eq:tradeoff}) is met with equality. This applies to higher number of users per set as well \cite{Berger99}.

The above conditions correspond to the ideal case where the rate of each source is fixed. But the rate allocation depends on the source grouping strategy, as well as the inter-cell interference. Even if it cannot be used directly in our framework, the theoretical rate-distortion performance still provides a few valuable lessons that we use in constructing our grouping methods. One is that the direct benefit of joint decoding is larger when sources are grouped such that the intra-group correlation is maximized. Additionally, the channel quality for sources in a group affect the final correlation gain. As an example, if a source is jointly decoded with sources that have very low data rates, the benefit achieved by use of correlation is small, even if correlation level is high among the group. 

Additionally, recall that we must also take the indirect effects of grouping into account. The indirect effects of grouping are difficult to predict, especially since the prediction requires knowledge of the load in nearby cells. In Figure \ref{fig:groupings} (a) and (b), we illustrate two source grouping methods, one which only considers the direct effects of SG and another that also consider the indirect effects by prioritizing the outer-cell, interference causing users.

We propose two methods for source grouping. These are simple, constructive solutions to an NP-hard problem that cannot be solved exhaustively in real time. Each of the two methods is appropriate for different problem settings. The first method is a distortion-based grouping with outer priority (Distortion OP), which can be used for static networks, where sources are not added or removed, and channel conditions can be assumed constant. The second method is a distance-based grouping with outer priority (Distance OP) which is an adaptive method appropriate for non-static networks. This method does not use the R-D region to find the grouping, and simply uses the fact that the correlation likely decreases with distance between sources. We describe each method in more detail below.

\emph{\textbf{Distortion-based grouping with OP}}: The system initially performs independent decoding until performance is stable. Then we pick a random source from N outer-most sources in the cell and compute its expected performance if it is paired with any of the other sources in the cell using conditions in (\ref{eq:ld}) and (\ref{eq:ldeq}). We choose the pairing that results in the lowest distortion and remove the pair from the set. We repeat the process until no node remains and we measure the resulting distortion in the cell. The process is repeated T times with different initializations, and we finally keep the grouping with lowest distortion.

\emph{\textbf{Distance based grouping with OP}}: We first pick a random source from the N outer-most sources in the cell and pair it with its closest neighbor and remove the pair from the set. We repeat the process until no nodes remain and calculate the sum of inter-group distances in the cell. We repeat the method T times and choose the source grouping that achieves the lowest sum of inter-group distances.

\begin{figure}[htb]
\begin{minipage}[b]{1.0\linewidth}
 \centering
 \centerline{\epsfig{figure=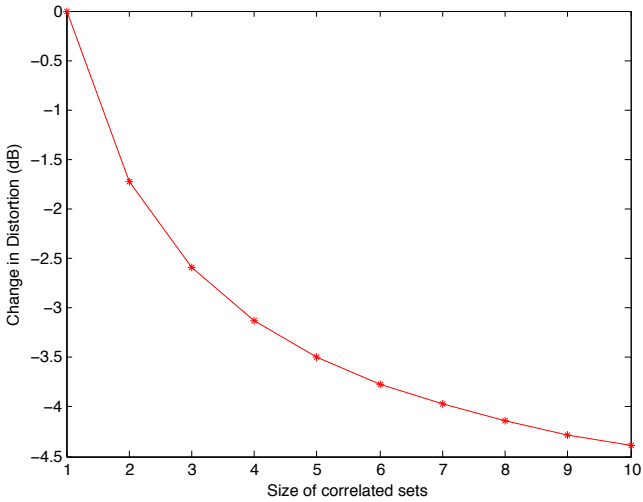,width=12cm, trim=1cm 9cm 1cm 8.5cm}}% L B R Top
 \caption{ Plot of \(\frac {\delta \Delta} N\) vs N, diminishing returns of increasing correlated set size on distortion.}
\label{fig:setsize}
\end{minipage}
\end{figure}

Finally, we discuss the benefits in source distortion when the size of the group increases. When grouping is performed on more than two sources, the overall benefit in log distortion can be approximated as:
\begin{align}
&\delta \Delta = \left[ h_2(\mathbf{S} ) - \sum_{X_i \in \mathbf{S}} h_2(Xi) \right ] & |\mathbf{S}| = N.
\end{align}
If all users have equal distortion levels, each users' distortion is decreased by \(\frac {\delta \Delta} N\) as a result of joint decoding. We plot this value for various set sizes, with entropy given by Eq. (\ref{eq:ent}), the covariance matrix given by Eq. (\ref{eq:corr}), and with distances between sources assumed equal. Figure \ref{fig:setsize} demonstrates the diminishing returns of set size. Additionally having large sets increases the Slepian-Wolf decoder complexity, as well as intra-cell scheduling complexity. Therefore we generally consider grouping of only a small number of sources, typically two or three nodes per set. 

When this step is performed, the scheduler of cell \emph{k} has an updated source grouping (i.e., \(\boldsymbol {G^*_k}\) in Eq. (\ref{eq:indepopt})). Given this value and \(\boldsymbol p^*\) found in the previous step the optimization problem of Eq. (\ref{eq:indepopt}) is simplified and the intra-cell scheduling unit can now find the channel allocation.

\subsection{Intra-Cell Scheduling}
\label{sec:intra}
In the previous phases of the three-step strategy we determined \(\boldsymbol p^*\) and \( \boldsymbol G^*\), i.e., power value and source grouping. In this last step we find \(\boldsymbol a^*\), namely the channel allocation to each user for the frames in the given scheduling period. 

We propose two solutions for this phase with different complexity and performance. For schedulers with low computational capacity, we propose the Distortion Proportionally Fair algorithm (D-PF). This solution is similar to the Proportionally Fair (PF) scheduling method \cite{Wengerter2005}; it is however modified to use distortion as utility instead of rate, and to take the correlation effects into account. D-PF scheduling can be performed at every frame. The second solution that we propose is called OPT and is appropriate for schedulers with higher computational capacity, more specifically, the ones that are capable of solving a linear programming scheduling problem at every scheduling period. OPT is a relaxed optimal assignment of channels, i.e., the relaxed version of Problem 2 given in (\ref{eq:indepopt}). This method is solvable in polynomial time. We present below both intra-cell scheduling algorithms in more detail.

\emph{\textbf{D-PF scheduling}}: With the common PF scheduling, the user that maximizes the following ratio is assigned channel \emph{c} at each frame \cite{Wengerter2005}:
\begin{align}
& i = \argmax_j \left(\frac{ R^{*}_{j,c}} {\overline{R}_j^\alpha} \right) \:\:\: \Rightarrow \:\:\: a^*_{i,c} = 1 \text{ and } a^*_{j,c} =0, \forall j \neq i. \nonumber
\end{align}
\(R^{*}_{i,c}\) is the possible rate achieved for user \emph{i} on channel \emph{c} is given by Eq. (\ref{eq:possrate}). \(\overline{R}_j\) is the exponentially averaged rate of user \emph{j} over the previous \(N_T\) frames, given as
\[\overline R_{j}(t-1) = (\frac {1} {N_T }). \sum_{c=1}^C R^*_{j,c}(t-1)  +  (1-\frac {1} {N_T }).\overline R_{j}(t-2). \]
The parameter \(\alpha\) is used to vary the trade-off between fairness and sum-rate maximization. The PF scheduling method is simple, yet performs almost the same as relaxed optimal channel allocation for independent sources, as we demonstrated in \cite{Wimob2011}. However, for correlated sources it is not sufficient to maximize the above ratio, since it ignores the effects of joint decoding. 

Therefore, we first define \(D^*_{i,c}\) as the possible distortion achieved by user \emph{i} if scheduled on channel \emph{c}, and  \(D^*_{j}\) as the possible distortion achieved by its correlated group members, i.e., for \(j \in \boldsymbol G \) and \(j \neq i\). In order to find these values, we must use conditions of Eqs (\ref{eq:ld}) and (\ref{eq:ldeq}) in Section \ref{sec:SG} for the group of correlated sources \(\boldsymbol {G}\) that \emph{i} belongs to. These conditions require the knowledge of the values of possible data rate that the source \emph{i} achieves if assigned channel \emph{c} in the next scheduling period, as well as the data rates of its correlated group members, which are therefore \emph{not} assigned channel \emph{c}. We assume that for every \emph{c}, \(R_{i}=R^{*}_{i,c}\), and \(R_{j} = 0\) for \(j \in \boldsymbol G \) and \(j \neq i\).
Given these possible data rates, then we can use the conditions in Eqs (\ref{eq:ld}) and (\ref{eq:ldeq}) to find \(D^*_{i,c}\) and \(D^*_{j}\). 

We finally propose the Distortion Proportionally Fair algorithm (D-PF) which uses the following condition for scheduling user \(l\) to channel \emph{c}:
\begin{align}
l = \argmin_i \left(\frac{ D^*_{i,c}} {\overline{D}_i^\alpha}. \prod_{j \in \boldsymbol G^*(i)}{\frac{D^*_j} {\overline{D}_j^\alpha}} \right)  \:\:\:\:\:\: \Rightarrow \:\:\:\:  a^*_{l,c} = 1 \text{ and } \:\:\:\: \nonumber \\
 a^*_{i,c} =0, \forall i \neq l \nonumber
\end{align}
where \(\overline{D}_j\) is the exponentially averaged distortion of user \emph{j} over the previous \(N_T\) frames, given as
\begin{align}
\overline D_{j}(t-1) = (\frac {1} {N_T }). D^*_{j}(t-1)  +  (1-\frac {1} {N_T }).\overline D_{j}(t-2) 
\label{eq:aveD}
\end{align}
And \(D^*_{j}(t-1)\) is found at the previous scheduling period after the channel assignment matrix \(\boldsymbol a^*\) is determined, using the conditions (\ref{eq:ld}) and (\ref{eq:ldeq}) with rates equal to \(R_i = \sum_{c=1}^C a^*_{i,c}.R^*_{i,c}\). The output of this method is then the matrix \(\boldsymbol a^*\), which are transmitted to the sources at every scheduling period. The above conditions can be readily generalized to more than two correlated sources per decoding group, with the conditions (\ref{eq:ld}) and (\ref{eq:ldeq}) replaced by the inequality set of Eq. (\ref{eq:rd}), solved for decreasing the maximum distortion. 

\emph{\textbf{OPT scheduling}}: Using this method, each cell finds the scheduling matrix \(\boldsymbol a^*\) by minimizing the maximum distortion in the cell, with \(\boldsymbol p^*\) and \( \boldsymbol G^*\) given in the previous steps. For this, we use Problem 2 in Eq. (\ref{eq:indepopt}) and relax it to become a linear programming problem. Namely, the values of  \(\boldsymbol a\) are relaxed to be real valued, allowing for time-sharing of each channel between users. Specifically, we perform OPT scheduling once every T frames and the real valued vector \(\boldsymbol {\tilde a}\) is rounded to achieve the corresponding resolution. The problem becomes the following for cell \emph{k}:
\begin{align}
\text{Problem 3:}\nonumber \\
\minimize_{\boldsymbol  {\tilde a} } \:\:\:  &\max_{i} (\Delta_{i}+\overline \Delta_{i}) \nonumber \\
&\sum_{s_i \in S} \sum_{c=1}^C \tilde a_{i,c} R^{*}_{i,c} \geq - { \frac {|S|} 2} \log_2 \left(2\pi e\right) + \sum_{s_i \in S} \Delta_{i}  \nonumber \\
& \:\:\:\:\:\:\:\: \:\:\:\: + h_2(\boldsymbol S |\boldsymbol G^j_{k} \backslash \boldsymbol S) ,  \:\:\:  \:\:\:\:\:\:\:\:\:\:\:\: \forall \boldsymbol S \subseteq \boldsymbol G^j_{k}, \forall j, \boldsymbol G^*_{k} \nonumber \\
&\sum_{i \in k} \tilde a_{i,c} = T , \:\:\: \:\:\:\:\:\:\:\: \:\:\:\:\:\:\:\:\:\:\:\: \:\:\:\: \:\:\:\:\:\:\:\:\:\:\:\: \:\:\:\:\:\:\:\:\:\:\:\:\:\:\:\:\:\:\:\: \:\:  \forall c \nonumber \\
& 0 \leq \tilde a_{i,c} \leq T, \:\:\:\:\:\:\:\: \:\:\:\:\:\:\:\:\:\:\:\:\:\:\:\:\:\:\:\:\:\:\:\: \:\:\:\:\:\:\:\:\:\:\:\: \:\:\:\:\:\:\:\: \:\: \forall i, \forall c. \nonumber
\end{align}
\(\overline \Delta_{i}\) is \( \log_{10} \overline D_{i}\), given by Eq. (\ref{eq:aveD}) and \(R^{*}_{i,c} \) is given by Eq. (\ref{eq:possrate}). This is a linear programming problem in \(\boldsymbol  {\tilde a}\), solvable in polynomial time using a method such as simplex algorithm \cite{vand}. After this problem is solved, the matrix  \(\boldsymbol  {a}\) is found by rounding \(\boldsymbol  {\tilde a}\) such that each \emph{c} is assigned to a single source in every frame in the following scheduling period. 
As with the case of D-PF, the matrix \(\boldsymbol  {a}\) is transmitted to users, determining the channel allocation in the following scheduling period. We compare these two scheduling methods in the simulations section below.

\section{Results}
\label{sec:sim}

We simulate a 19-cell hexagonal 2-D cellular network, with wrap around in order to avoid boundary inconsistencies. In each instance of the problem, 18 sources are placed randomly with uniform distribution in every cell. Simulation parameters are given in Table \ref{tab:param}, and are chosen in accordance with the micro test case in LTE \cite{3GPP2006}. The observations at the sources are modeled as joint Gaussian random variables, and an exponential distance based correlation model is assumed as in Eq. \ref{eq:corr}. We use the Cumulative Density Function (CDF) of the distortion values and rates achieved by all sources in order to demonstrate the performance of different resource allocation algorithms. This metric illustrates the performance of all the sources in every cell, from the highest performing ones to the most disadvantaged. Since the aim in this work is to minimize the maximum distortion achieved in the network, the parameter we look for is the performance of 5 percentile worst performing users (5 percentile rate and 95 percentile distortion). We begin by comparing the performance of various methods for each step of the three-step algorithm, in order to justify the different choices we have made in the proposed scheme. Later we will compare the overall performance gain of utilizing correlation in resource allocation using our proposed scheme versus independent allocation of resources. Finally, we discuss the convergence of the proposed algorithm.

\begin{figure*}[!t]
\centerline{\subfigure[Distortion of various inter-cell resource management methods.]{\includegraphics[width=9.8cm,trim=2cm 8cm 3.5cm 8.5cm]{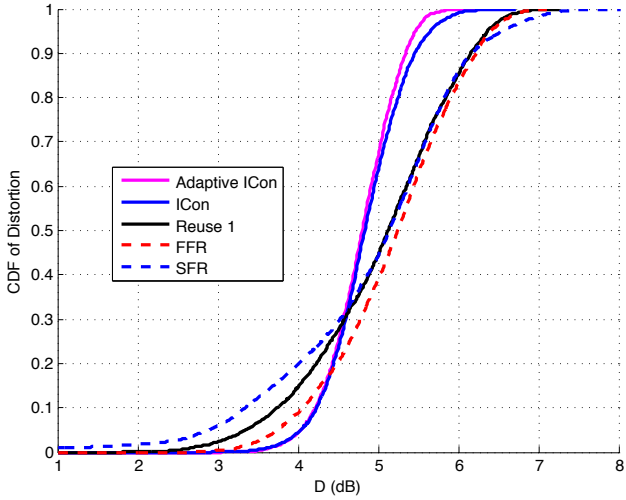} 
\label{fig:interD}}
{\subfigure[Rate of various inter-cell resource management methods.]{\includegraphics[width=9.8cm,trim=3.5cm 8cm 2cm 8.5cm]{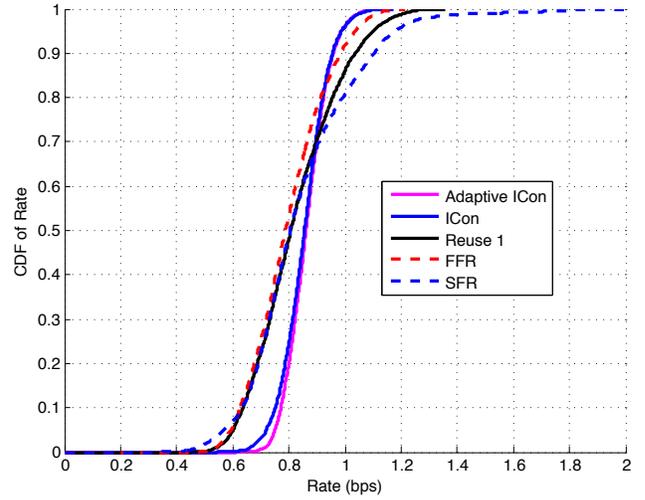}
\label{fig:interR}}}}
\caption{Inter-cell scheduling methods.}
\end{figure*}

\begin{figure}[h!]
\begin{minipage}[b]{1.0\linewidth}
 \centering
 \centerline{\epsfig{figure=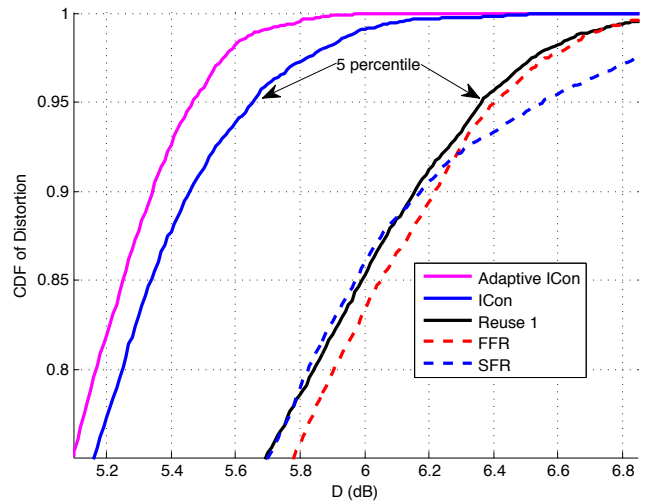,width=12cm, trim=1cm 8.5cm 1cm 8.5cm}}% L B R Top
 \caption{Closeup of Distortion CDF of inter-cell resource management methods.}
\label{fig:closeup}
\end{minipage}
\end{figure}

In order to compare the inter-cell resource management methods, we simulate FFR, SFR, Reuse 1, static ICon, and Adaptive ICon schemes. In this step we do not use correlation, since we would like to isolate the effects of inter-cell resource management methods. We compare the CDFs of distortions and rates achieved by sources, as shown in Figures \ref{fig:interD} and \ref{fig:interR}, with 5 percentile distortion details shown in Figure \ref{fig:closeup}. We observe that using our static ICon inter-cell interference management method compared with Reuse 1, FFR, and SFR, the 95 percentile distortion is decreased by 0.75 dB. FFR, SFR and Reuse 1 perform similarly, with Reuse 1 having a slight advantage, as we expected. With little communication between base-stations of neighboring cells and Adaptive ICon, this gain is increased to 1 dB.

\begin{figure*}[htb]
\centerline{\subfigure[Comparison of the grouping methods (2 sources per group).]{\includegraphics[width=9.8cm,trim=2cm 8cm 3.5cm 8.5cm]{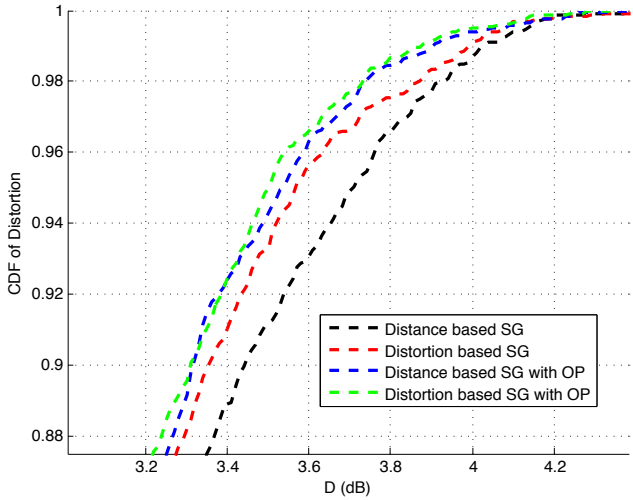} 
\label{fig:SGop}}
{\subfigure[Influence of the size of the groups in the resource allocation scheme.]{\includegraphics[width=9.8cm,trim=3.5cm 8cm 2cm 8.5cm]{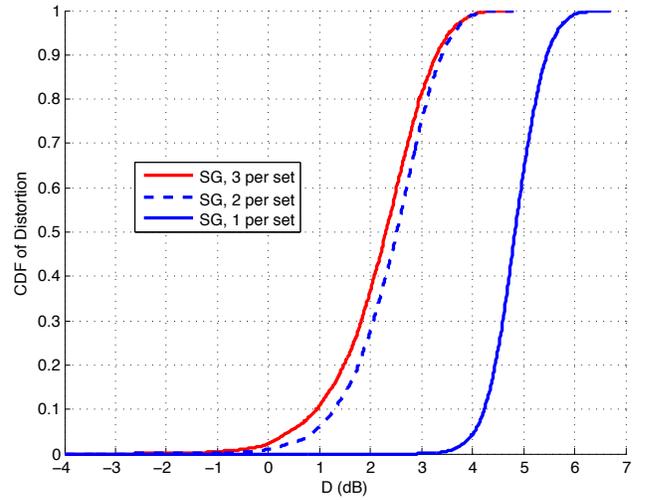}
\label{fig:SGsize}}}}
\caption{Source grouping methods.}
\label{fig:SG}
\end{figure*}

We now add correlation to the resource allocation method in order to compare source grouping methods in Figure \ref{fig:SG}. We use D-PF for intra-cell allocation for all methods in this part of the analysis. We first compare in Figure \ref{fig:SGop} the two grouping methods proposed in Section \ref{sec:SG}, namely distance OP and distortion OP for groups of two sources. We also show the effects of using outer priority in the grouping algorithm. We find that distortion OP performs slightly better than distance OP, however distortion OP cannot be used adaptively. We also compare the effects of the group size in Figure \ref{fig:SGsize}. Increasing set size from one user per group (i.e., independent decoding and independent resource allocation) to two decreases the 95 percentile distortion by 2 dB (37\%). However, there is almost no difference in performance when we increase the group size from two to three sources, which is expected as explained in Section \ref{sec:SG}. 

\begin{figure}[htb]
\begin{minipage}[b]{1.0\linewidth}
 \centering
 \centerline{\epsfig{figure=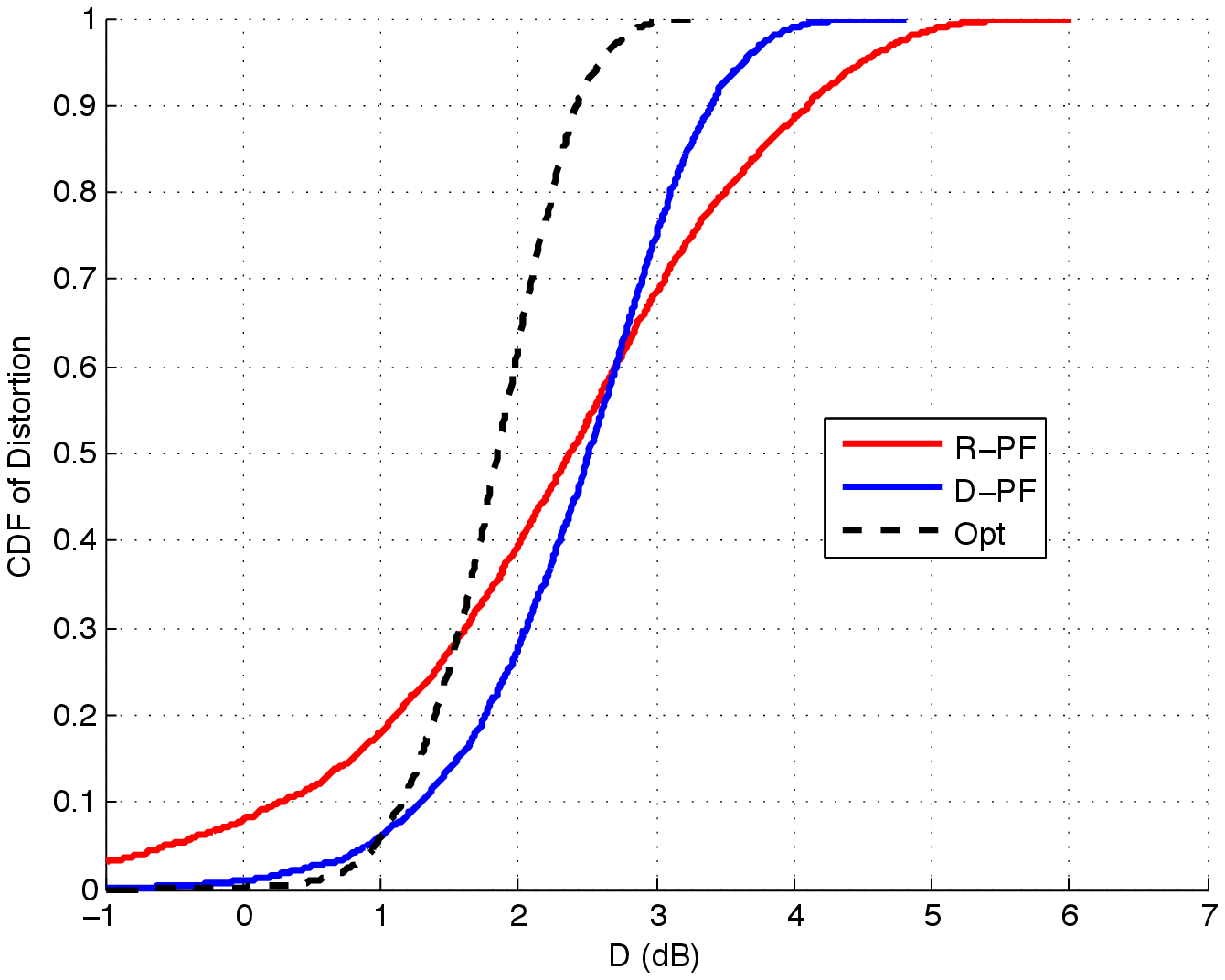,width=12cm, trim=1cm 8.5cm 1cm 8.5cm}}% L B R Top
 \caption{Comparison of CDF of Distortion in all cells, in D-PF, R-PF, and OPT scheduling.}
\label{fig:intraD}
\end{minipage}
\end{figure}

We then compare the intra-cell scheduling methods, namely, PF, D-PF, and OPT, with inter-cell method given by static ICon and source grouping of 2 per set Distance OP for all comparisons. The results are shown in Figure \ref{fig:intraD}. The PF curve in this figure demonstrates the case where correlation is used in compression and joint decoding, but not in resource allocation. Comparison of the PF curve with the independent decoding and resource allocation case, demonstrated in Figure \ref{fig:SGsize} in the case of 1 source per group, highlights the benefits of using correlation in compression, achieving 1.25 dB (or 25\%) decrease in 95 percentile distortion. On the other hand, comparison of D-PF with PF highlights the added benefits of correlation-aware resource allocation, achieving an additional 0.75 dB decrease in 95 percentile distortion, adding up to 37\% loss in distortion. Using OPT scheduling instead of D-PF increases this distortion loss further, adding up to 1.75 dB over PF scheduling, for a total of 50\% loss over the independent method. This is a large gain, which is feasible if the scheduler has the computational capacity to perform OPT, as discussed in Section \ref{sec:intra}.

\begin{figure}[htb]
\begin{minipage}[b]{1.0\linewidth}
 \centering
 \centerline{\epsfig{figure=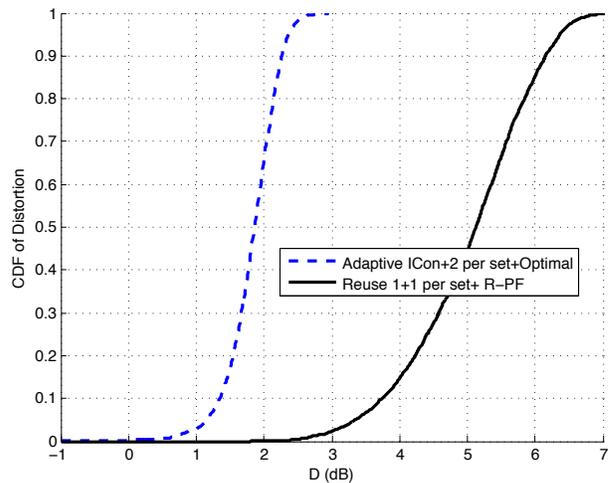,width=8cm,trim=4cm 8cm 4cm 8.5cm}}% L B R Top
 \caption{Comparison of our resource allocation algorithm with a baseline scheme.}
\label{fig:final}
\end{minipage}
\end{figure}

Finally we show in Figure \ref{fig:final} the overall performance of the three-step strategy, with Adaptive ICon, Distance OP source grouping (two users per set), and optimal intra-cell scheduling. We compare to a baseline method, namely Reuse 1 without spatial correlation and PF scheduling, common in cellular networks. Overall, we demonstrate that this method achieves a large improvement with almost a 4 dB (60\%) decrease in distortion for 5 percentile users.

\begin{table}[htb]
\renewcommand{\arraystretch}{1.3}
\caption{Simulation Parameters}
\label{tab:param}
\centering
\begin{tabular}{|l||c|}
\hline
Parameter & Values\\
\hline
Number of cells & 19, with wrap around\\
%Users per cell for static simulations & 20, uniformly distributed\\
Users per cell& 18, uniformly distributed\\
Site to site distance  (m) & 130\\
Bandwidth & 10 Hz \\
Number of subbands & 63 \\
Max Power per user per band &  250mW/\(63 \times 10^{-6}\) \\
Path loss model (dB) & \(30 \log_{10} R\), R in (m). \\
Fractional power control & \(\alpha = 1\), \(\Gamma\) differs.\\
Interference over Thermal (IoT) & ICon=10, Reuse1=13 (dB)\\
PF scheduling parameters& a = 3.5\\
FFR reuse 1 band/total band& \(\eta=45/63\)\\
SFR parameters& \(p_l/p_h = 1/10\), \(\gamma = 6\) dB\\
PF Scheduling period & 1 frame \\
Optimal Scheduling period &  10 frames \\
Step sizes for adaptive ICon & \(\alpha = \beta = 0.2\)\\
Gaussian sources mean and var&\(\sigma^2\)=10, mean = 0 \\
Correlation parameter&\(\theta\)= 100\\
\hline
\end{tabular}
\end{table}

\section{Conclusion} 
In this work we consider the problem of resource allocation between spatially correlated sources in FDMA multi-cell networks. This is an NP-hard problem for which exhaustive solutions are not computationally feasible in practice. We propose a cross-layer solution that performs effective but suboptimal resource allocation in three simple steps. The design parameters are power per channel per source, and the grouping of sources to be decoded jointly. We determine the power per channel per user in the inter-cell resource management step, the grouping of sources for joint decoding in the source grouping step and the channel allocation to each source in the intra-cell scheduling step. We evaluate our design choices in the simulations by comparing various methods for each step of the algorithm. 

Overall, the performance gain of our proposed scheme over baseline independent scheduling methods is a 4 dB decrease in distortion of the worst performing sources. Additionally, we show that while the benefit of using correlation in joint decoding is 25\% in distortion, using a simple correlation-aware resource allocation increases the loss to 37\%. This confirms the benefit of considering the correlation of sources directly in the resource allocation problem. We plan to extend this study to the transmission of different sources.

\bibliographystyle{IEEEbib}
\bibliography{correlatedmacv2}
%\bibliography{corrsources,strings,dvc,resourcealloc,gameth,WSNs,wimob}

\end{document}